\documentclass[twocolumn,pra,showpacs]{revtex4}
\usepackage{amssymb}
\usepackage{amsmath}
\usepackage{graphicx}
\usepackage{epsfig}
\usepackage{subfigure}
\usepackage{amsfonts}

\begin{document}

\title{Decoherence-free dynamics of quantum discord for two correlated qubits in finite-temperature reservoirs}
\author{Lan Xu$^{1,2}$, Ji-Bing Yuan$^{1}$, Qing-Shou Tan$^{1}$, Lan Zhou$^{1}$, and Le-Man Kuang$^{1}$}
\thanks{Author to whom any correspondence should be
addressed} \email{lmkuang@hunnu.edu.cn} \affiliation{$^{1}$ Key
Laboratory of Low-Dimensional Quantum Structures and Quantum
Control of Ministry of Education, and Department of Physics, Hunan
Normal University, Changsha 410081, China}
\affiliation{$^{2}$Department of Mathematics and Science, Hunan
First Normal University, Changsha 410205, China}

\begin{abstract}
We investigate decoherence-free evolution (DFE) of quantum discord
(QD) for two initially-correlated qubits in two finite-temperature
reservoirs using an exactly solvable model. We study QD dynamics
of the two qubits coupled to two independent Ohmic reservoirs when
the two qubits are initially prepared in $X$-type quantum states.
It is found that reservoir temperature significantly affects the
DFE dynamics. We show that it is possible to control the DFE and
to prolong the DFE time by choosing suitable parameters of the
two-qubit system and reservoirs.


\end{abstract}

\maketitle
\narrowtext

\section{\label{Sec:1}INTRODUCTION}

It is well known that quantum entanglement \cite
{RMP73(09)00565,RMP77(05)00633,RMP73(01)00357,physrep04,physrep02,physrep05,RMP81(09)00865},
is a distinctive quantum feature of quantum correlations, but not
the only one. In order to capture nonclassical correlations,
several measures of correlations have been proposed in the
literature \cite
{JPA34,oh3PRL89,GPWPRA72,SLPRA77,VedPRL104,DLZPRL101,zerekdMD}.
Among them, quantum discord (QD) has received considerable
attention \cite
{PRL105(10)020503,PRL100(08)090502,JPA41(08)205301,Lidar102,DGPRA79,Yuan43},
which quantifies the quantumness of correlations between two
partitions in a composite state \cite{zurekPRL88}. The QD, which
measures general nonclassical correlations including entanglement,
is defined as the mismatch between two quantum analogues of
classically equivalent expression of the mutual information. For
pure entangled states, all nonclassical correlations which can be
characterized by the QD are identified as entanglement. For mixed
states, the entanglement can not totally describe nonclassical
correlation. Even for some separable states, although there is no
entanglement between these two parts, the QD is nonzero, which
indicates the presence of nonclassical correlations. And compared
to classical counterparts, such separable states can also improve
performance in some computational tasks \cite{Datta100,LBAW101}.
In fact, Nonclassical correlation described by the QD can be
considered as a more universal quantum resource than quantum
entanglement in some sense, and the QD offers new prospects for
quantum information processing.

Any realistic quantum system will inevitably interact with the
surrounding environment, which causes the rapid destruction of
crucial quantum properties. Therefore, besides the
characterization and quantification of correlations, an
interesting and crucial issue is the behavior of correlations
under decoherence. The QD dynamics have been widely studied by
using models where qubits interacts in zero-temperature reservoirs
\cite{Wang2009,MWPRA81,Yuan43,Piilo2010,VedPRA80,cel,xu1,xu2}.
Moreover, QD is more robust than the entanglement against
decoherence \cite{WSPRA80,FWPRA81,FAPRA81,MWPRA81}. Also, unlike
the entanglement, which exhibits sudden death
~\cite{Yu2004,Bellomo2007,Choi2007,opez2008,Qasimi2008,Maniscalco2008},
the QD displays the sudden-transition phenomenon from a
decoherence-free-evolution (DFE) to a decoherence-evolution regime
in the dynamic evolution \cite{Piilo2010,VedPRA80,cel}, and the
inital QD remains unchanged in the DFE regime. The This
sudden-change phenomenon have been demonstrated in recent
experiments \cite{xu1,xu2}. It is of significant interest to
prolong the DFE time of the QD in quantum information science
since quantum information processing favors the long DFE time of
the QD in the dynamic evolution. In this paper, we study the
possibility of prolonging the DFE time of the QD by investigating
the DFE dynamics of initially-correlated two qubits in two
finite-temperature reservoirs. We shall show that the DFE time of
the QD  can be controlled by choosing initial-state parameters of
the two qubits, reservoir parameters, and qubit-reservoir
interactions.

This paper is organized as follows: In Sec.~\ref{Sec:2}, we
present our physical model and study its solution. In
Sec.~\ref{Sec:3}, we study the DFE dynamics of the QD to indicate
the possibility of prolong the DFE time of the QD by investigating
QD dynamical behaviors of the two qubits in finite-temperature
reservoirs for certain initially prepared $X$-type states. The
effects of the temperature, the initial-state parameters, the
system-reservoir coupling on QD are studied in Sec.~\ref{Sec:4}.
Finally, we conclude this work in the last section.

\section{\label{Sec:2} the model and its solution}

\begin{figure}[tbp]
\includegraphics[bb=92 527 395 638, width=3.0 in]{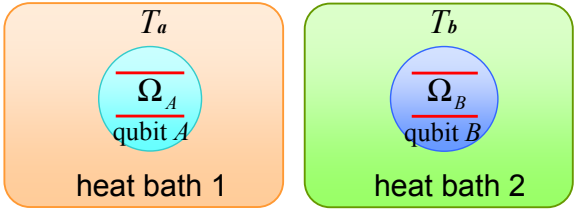}
\caption{(Color online) Schematic of a pair of noninteracting
qubits A and B with energy separations $\Omega _{A}$ and $\Omega
_{B}$ under the influence of two independent environments with
temperature $T_{a}$ and $T_{b}$.} \label{fig1.eps}
\end{figure}

The total system we consider is shown in Fig.~\ref{fig1.eps},
including a pair of noninteracting qubits A and B in two
independent heat baths, respectively. The energy separation
between the excited state $\left\vert e\right\rangle $ and ground
state $\left\vert g\right\rangle $ of qubits is denoted by $\Omega
_{i} $ ($i=A,B$). Atoms A and B are coupled individually to its
thermal bath with temperature $T_{a}$ and $T_{b}$ respectively.
Here, each bath is
modelled as an infinite number of harmonic oscillators with frequencies $%
\omega _{ak}$ and $\omega _{bk}$, which couple to the relevant
electronic degrees of freedom of  $i$-th qubit via the coupling
constants $g_{ik}$. The Hamiltonian of the total system including
the two qubits and the environment is composed of three parts
\begin{equation}
H=H_{S}+H_{B}+H_{SE}, \label{qd2-1}
\end{equation}%
where Hamiltonians of the system and baths are given by
\begin{equation}
H_{S}=\frac{\Omega _{A}}{2}{\sigma }_{A}^{z}+\frac{\Omega _{B}}{2}{\sigma }%
_{B}^{z},  \label{qd2-2}
\end{equation}
\begin{equation}
H_{B}=\sum_{k}\hbar \left( \omega _{ak}{a}_{k}^{\dag }{a}_{k}+\omega _{bk}{b}%
_{k}^{\dag }{b}_{k}\right),  \label{qd2-3}
\end{equation}
where ${\sigma }_{i}^{z}$ is the standard diagonal Pauli matrix,
and ${a}_{k}^{\dag }$(${b}_{k}^{\dag }$) and ${a}_{k}
$(${b}_{k}$) are the creation and annihilation operators for a oscillator in $k$%
the mode, obeying the bosonic commutation relation. The
system-bath interaction Hamiltonian is given by
\begin{equation}
H_{SE}=\sum_{k}\hbar \left( {\sigma }_{A}^{z}g_{Ak}{a}_{k}+{\sigma
} _{B}^{z}g_{Bk}{b}_{k}+{\texttt{h.c.}}\right).  \label{qd2-4}
\end{equation}

The bilinear interaction between the system and the bath in $H_{SB}$
indicates that the atomic inverse operator ${\sigma }_{i}^{z}$ commutates
with Hamiltonian $H$, i.e. $\left[ \sigma _{i}^{z},H\right] =0$. In the
rotating frame with respect to Hamiltonian $H_{S}+H_{B}$, the interaction
Hamiltonian reads%
\begin{equation}
\tilde{H}\left( t\right) =\sum_{k}\hbar \left( g_{Ak}e^{-i\omega _{ak}t}{%
\sigma }_{A}^{z}{a}_{k}+g_{Bk}e^{-i\omega _{bk}t}{\sigma }_{B}^{z}{b}_{k}+{%
\texttt{h.c.}}\right) ,  \label{qd2-5}
\end{equation}%
Equation (\ref{qd2-5}) indicates that the state of the environment
will be sensitive to the values of $\sigma _{i}^{z}$. The
commutation $\left[ H_{S},H_{SE}\right] =0$ suggests that no energy
exchanges between qubit and bath, i.e., energy of the system $S$ is
conservative \cite {BBcontrol,Kuang1999}. Therefore, the model
describes a purely decohering
mechanism. The evolution operator generated by the effective Hamiltonian (%
\ref{qd2-5}) is given by  $U\left( t\right) =U_{a}\left( t\right)
U_{b}\left(
t\right) $ with%
\begin{equation}
U_{i}(t)=\exp \left\{ \frac{\sigma _{i}^{z}}{2}\sum_{k}\left[
X_{k}\xi _{Xk}^{\ast }\left( t\right) -\texttt{h.c.}\right]
\right\} \label{qd2-6}
\end{equation}%
where  $\xi _{Xk}\left( t\right) =\frac{2g_{Xk}}{\omega
_{Xk}}\left( 1-e^{i\omega _{Xk} t}\right) $. Here, $X=a(b)$ when
$i=A(B)$. The analytical solution of this model allows us to study
behaviors of these correlations under the action of decoherence
without any approximations.

In order to show how decoherence affects the correlations in this
two-qubit composite system, we assume that the initial state of
the two baths is a thermal state denoted by the following density
operator
\begin{equation}
\rho _{E}\left( 0\right) =\prod_{X,k}\left( 1-e^{-\beta _{X}\hbar
\omega _{Xk}}\right) e^{-\beta_{X} \hbar \omega _{Xk}X_{k}^{\dag
}X_{k}}  \label{qd2-7}
\end{equation}%
where $\beta _{X}=\left( k_{B}T_{X}\right) ^{-1}$, and the initial
state of the two qubits is denoted by the density operator $\rho
_{S}\left( 0\right) $, the
total system is assumed as a product state of these two initial states $%
\rho \left( 0\right) =\rho _{S}\left( 0\right) \otimes \rho
_{E}\left( 0\right) $. By tracing out over the state of the
environment, we can obtain the quantum state of the two qubits at
any time
\begin{equation}
\rho _{S}\left( t\right) =\texttt{Tr}_{E}\left[ U^{\dag }\left(
t\right) \rho \left( 0\right) U\left( t\right) \right].
\label{qd2-8}
\end{equation}

In the Hilbert space spanned by the two-qubit product state basis
$\left\{\left\vert gg\right\rangle ,\left\vert ge\right\rangle
,\left\vert
eg\right\rangle ,\left\vert ee\right\rangle \right\} $, the density operator $%
\rho _{S}\left( t\right) $ loses its off-diagonal terms as a result of the
interaction with the environment $E$. The elements of the density operator $%
\rho _{S}\left( t\right) $ can be exactly obtained as
\begin{eqnarray}
\rho _{(l_{A},l_{B})(j_{A},j_{B})}(t) &=&\rho
_{(l_{A},l_{B})(j_{A},j_{B})}(0)  \notag \\
&&\times \exp {[(\delta _{l_{A},j_{A}}-1)\Gamma _{A}\left(
t\right) ]}
\notag \\
&&\times \exp {[(\delta _{l_{B},j_{B}}-1)\Gamma _{B}\left(
t\right) }], \label{qd2-9}
\end{eqnarray}%
where $l,j=e,g$ and ${\delta _{l,j}}$ is the Kronecker delta
function. Here, we note that no approximation is employed.
Equation (\ref{qd2-9}) shows that diagonal terms of the reduced
density matrix $\rho _{S}\left( t\right) $ remain in the initial
value, however, the off-diagonal terms vary with the time
evolution. The two time-dependent decohering factors are defined
by
\begin{equation}
\Gamma _{i}\left(t,0\right) =\sum_{X,k}\frac{\left\vert \xi
_{X,k}\left( t\right) \right\vert ^{2}}{2}\coth \left(\frac{\hbar
\beta _{X}\omega _{Xk}}{2}\right),  \label{qd2-10}
\end{equation}%
which are two real-value functions completely to characterize the
two-qubit dynamics in the decohering process.

Since the states in the reservoir are very dense (continuum), we can
take the continuum limit to convert the summation over $k$ into an
integral with respect to $\omega _{X}$. Then the properties of
environment are described
by the reservoir spectral density%
\begin{equation}
J\left( \omega _{X}\right) =\sum_{X,k}\delta (\omega _{X}-\omega
_{X,k})\left\vert g_{X,k}\right\vert ^{2}.  \label{qd2-11}
\end{equation}%
With these replacements, the two decohering factors in Eq. (10)
become
\begin{eqnarray}
\Gamma _{i}\left( t\right) &=&8\int_0^{\infty }d\omega
_{X}\frac{J\left( \omega _{X}\right)}{\omega^2_{X}} \coth
\left(\frac{\hbar \beta _{X}\omega _{X}}{2}\right)
\label{qd2-12} \\
&&\times \sin^2\left(\frac{\omega _{X}t}{2}\right). \notag
\end{eqnarray}%

\section{\label{Sec:3}DFE dynamics of the QD for two initially
correlated qubits}

In this section we study DFE dynamics of the QD for the two
initially correlated qubits in the two independent
finite-temperature reservoirs by investigating the QD dynamics
under decoherence. The total correlations between two qubits A and
B described by a bipartite quantum state $\rho _{AB}=\rho _{S}$
are generally measured by quantum mutual information
\begin{equation}
\mathcal{I}\left( \rho _{S}\right) =S\left( \rho _{A}\right) +S\left( \rho
_{B}\right) -S\left( \rho _{S}\right) ,  \label{qd3-1}
\end{equation}%
where $S\left( \rho \right) =-\text{Tr}(\rho \log \rho )$ is the
von Neumann entropy of density matrix $\rho $, $\rho _{A}\left(
\rho _{B}\right) $ is the reduced density operators for subsystem
A(B). Quantum mutual information contains quantum correlation
$\mathcal{D}\left( \rho _{S}\right) $
and classical correlation $\mathcal{C}\left( \rho _{S}\right) $\cite%
{JPA34,VedPRL90,VedPRA80}. Quantum correlation $\mathcal{D}\left(
\rho _{S}\right) $ can be quantified by the QD \cite%
{zerekdMD,zurekPRL88}, which is defined by the discrepancy between
quantum versions of two classically equivalent expressions for
mutual information
\begin{equation}
\mathcal{D}\left( \rho _{S}\right) =\mathcal{I}\left( \rho _{S}\right) -%
\mathcal{C}\left( \rho _{S}\right) \text{,}  \label{qd3-2}
\end{equation}
where the classical correlation $\mathcal{C}\left( \rho
_{S}\right) $ is defined as the maximum information about one subsystem $%
\rho _{i}$, which depends on the type of measurement performed on
the other subsystem. For a local projective measurement $\Pi _{k}$
performed on the subsystem $B$ with a given outcome $k$, we denote
\begin{equation}
p_{k}=\text{Tr}_{AB}[(I_{A}\otimes \Pi _{k})\rho _{S}(I_{A}\otimes \Pi
_{k})].  \label{qd3-3}
\end{equation}%
as the probability, where $I_{A}$ is the identity operator for the subsystem
$A$. Then the classical correlation reads%
\begin{eqnarray}
\mathcal{C}(\rho _{S}) &=&\max_{\{\Pi _{k}\}}\left[ S(\rho
_{A})-\sum_{k}p_{k}S(\rho _{A}^{(k)})\right]  \label{qd3-4} \\
&=&S(\rho _{A})-\min_{\{\Pi _{k}\}}\left[ \sum_{k}p_{k}S(\rho _{A}^{(k)})%
\right] ,  \notag
\end{eqnarray}%
where the maximum is taken over the complete set of orthogonal projectors $%
\left\{ \Pi _{k}\right\} $ and
\begin{equation}
\rho _{A}^{(k)}=\frac{1}{p_{k}}\text{Tr}_{B}\left[ \left( I_{A}\otimes \Pi
_{k}\right) \rho (I_{A}\otimes \Pi _{k})\right]  \label{qd3-5}
\end{equation}%
is the reduced density matrix of subsystem $A$ after obtaining the
measurement outcome $k$, which is normalized.

Let us first find the analytical expression for the density
operator of two qubits at time $t$. We assume the two qubits are
initial in a class of states with maximally mixed marginals, which
is described by the $X$-structured density operator
\begin{equation}
\rho _{S}\left( 0\right) =\frac{1}{4}\left( I_{AB}+\underset{j=1}{%
\overset{3}{\sum }}c_{j}\sigma _{A}^{j}\otimes \ \sigma _{B}^{j}\right) .
\label{qd3-6}
\end{equation}%
Here, the coefficients $c_{j}$ with real constants satisfy the
condition that $\rho _{S}(0)$ is positive and normalized, and
$I_{AB}$ is the identity operator of the two qubits. Obviously,
such states $\rho _{S}\left(0\right)$ are general enough to
include states such as the Werner states and the Bell states.

Using the results obtained in Eq.(\ref{qd2-9}), the density matrix
of two qubits at time $t$ has the following analytical expression
\begin{equation}
\rho _{S}\left( t\right) =\frac{1}{4}\left(
\begin{array}{cccc}
1+c_{3} & 0 & 0 & \tilde{\alpha}^{\ast } \\
0 & 1-c_{3} & \tilde{\gamma }^{\ast } & 0 \\
0 & \tilde{\gamma } & 1-c_{3} & 0 \\
\tilde{\alpha} & 0 & 0 & 1+c_{3}%
\end{array}%
\right) ,  \label{qd3-7}
\end{equation}%
where $\tilde{\alpha}=\alpha(t) e^{-i(\Omega _{A}+\Omega _{B})t}$ and $\tilde{\gamma }%
=\gamma(t) e^{i(\Omega _{B}-\Omega _{A})t}$  with,
\begin{subequations}
\label{qd3-8}
\begin{eqnarray}
\alpha (t)&=&(c_{1}-c_{2})D_{A}(t)D_{B}(t), \\
\gamma (t)&=&(c_{1}+c_{2})D_{A}(t)D_{B}(t),
\end{eqnarray}%
where $D_{i}(t)=\exp \left[ -\Gamma _{i}\left( t\right) \right]$
are decohering functions. Since the reduced density matrix $\rho
_{i}\left( t\right) =I_{i}/2$ of subsystem $i=A,B$, and the
eigenvalues of the two-qubit density operator $\rho _{S}\left(
t\right) $ can be exactly calculated from Eq.(\ref{qd3-7}) with
the following expressions
\end{subequations}
\begin{subequations}
\label{qd3-9}
\begin{eqnarray}
\lambda _{1,2} &=&\frac{1}{4}\left( 1+c_{3}\mp \alpha \right) \\
\lambda _{3,4} &=&\frac{1}{4}\left( 1-c_{3}\mp \gamma \right),
\end{eqnarray}%
quantum mutual information can be written as
\end{subequations}
\begin{equation}
\mathcal{I}\left( \rho _{S}\right) \ =2+{\sum_{i=1}^{4}}\lambda _{i}\log
\lambda _{i}.  \label{qd3-10}
\end{equation}

Now, we turn to calculating the classical correlation $\mathcal{C}(\rho _{S})$ defined in Eq. (%
\ref{qd3-4}). In the Hilbert space spanned by $\left\{ \left\vert
e\right\rangle ,\left\vert g\right\rangle \right\} $, any two
orthogonal
states $\left\vert \mathbf{0}\right\rangle $ and $\left\vert \mathbf{1}%
\right\rangle $ can be represented as a unitary vector on the Bloch sphere
\begin{subequations}
\label{qd3-11}
\begin{align}
\left\vert \mathbf{0}\right\rangle & =\cos \theta \left\vert g\right\rangle
+e^{i\phi }\sin \theta \left\vert e\right\rangle , \\
\left\vert \mathbf{1}\right\rangle & =e^{-i\phi }\sin \theta \left\vert
g\right\rangle -\cos \theta \left\vert e\right\rangle ,
\end{align}%
with $0\leq \theta \leq \pi /2$ and $0\leq \phi \leq 2\pi $. Therefore, for
a local measurement performed on the subsystem $B$, a complete set of
orthogonal projectors contains two elements $\Pi _{\mathbf{0}}=\left\vert
\mathbf{0}\right\rangle \left\langle \mathbf{0}\right\vert $ and $\Pi _{%
\mathbf{1}}=\left\vert \mathbf{1}\right\rangle \left\langle \mathbf{1}%
\right\vert $ with the probability $p_{\mathbf{0}}=p_{%
\mathbf{1}}=1/2$. After the two project measurements the reduced
density operator of subsystem $A$ with an outcome $k$ reads
\end{subequations}
\begin{equation}
\rho _{A}^{\left( k\right) }=\frac{1}{4}\left(
\begin{array}{cc}
2(1-c_{3}\cos (2\theta )) & \left( -1\right) ^{k}\epsilon \sin (2\theta ) \\
\left( -1\right) ^{k}\epsilon ^{\ast }\sin (2\theta ) & 2(1+c_{3}\cos
(2\theta ))%
\end{array}%
\right)  \label{qd3-12}
\end{equation}%
where we have introduced the following parameter
\begin{equation}
\epsilon =\alpha e^{i\left[ (\Omega _{A}+\Omega _{B})t-\phi \right] }+\gamma
e^{i\left[ (\Omega _{A}-\Omega _{B})t+\phi \right] }\text{.}  \label{qd3-13}
\end{equation}%

Making use of Eq. (24), it is straightforward to calculate the
classical correlation with the following expression \cite{Yuan43}
\begin{equation}
\mathcal{C}\left( \rho _{S}(t)\right) =\overset{2}{\underset{n=1}{\sum }}%
\frac{1+(-1)^{n}\chi (t)}{2}\log _{2}\left[ 1+(-1)^{n}\chi
(t)\right], \label{qd3-15}
\end{equation}
where
\begin{equation}
\chi (t)=\max \left[ \left\vert c_{3}\right\vert ,(|\alpha(t)
|+|\gamma(t) |)/2\right],  \label{eq27}
\end{equation}
which depends on the relation between the coefficients $c_{j}$ of
the initial state in Eq.(\ref{qd3-6}) and the dynamic parameters
defined by Eq. (22). From Eq. (27) we can see that:  (i) If $
|c_{3}|>(|\alpha |+|\gamma |)/2$, we have $\chi (t)=\left\vert
c_{3}\right\vert $; (ii) If $|c_{3}|<(|\alpha |+|\gamma |)/2$, we
have $\chi (t)=(|\alpha |+|\gamma |)/2$.

Then the QD between the two  qubits  can be written as
\begin{equation}
\mathcal{D}(\rho_{S}(t))=2+\sum_{i=1}^{4}\lambda _{i}\log
_{2}\lambda _{i}-\mathcal{C}(\rho _{S}(t)). \label{qd3-16}
\end{equation}
In order to clearly understand the QD dynamics of two qubits in
two independent reservoirs, we consider the case  of  $c_{1}=1$,
$c_{2}=-c_{3}$, and $\ \left\vert c_{3}\right\vert <1$. In this
case, from Eq. (\ref{eq27}) we can obtain
\begin{equation}
\chi(t)=\max [|c_{3}|, D_{A}(t)D_{B}(t)],  \label{qd3-17}
\end{equation}
where the two decohring functions are given by
\begin{equation}
D_i(t)=(1+\omega _{ic}^{2}t^{2})^{-\frac{\eta_i }{2}}\prod_{n=1}^{\infty }\left[1+%
\frac{\omega _{ic}^{2}t^{2}}{(1+\hbar \beta_i \omega
_{ic}n)^{2}}\right]^{-\eta_i }, \label{qd3-17}
\end{equation}
where $i=(A,B)$ and we assume the two reservoirs have the Ohmic
spectral density \cite{LegRMP59} given by
\begin{equation}
J(\omega_i)=\eta_i\omega_i\exp \left(\frac{-\omega_i}{\omega
_{ic}}\right). \label{qd3-18}
\end{equation}
where $\eta_i > 0$ is the system-reservoir coupling constant, and
$\omega _{ic}$ is the high-frequency cut-off.

Taking into account the decohring functions $D_A(t)$ and $D_B(t)$
decay in the time evolution, making use of Eqs. (21), (26), (28),
and (29), we can find the critic time denoted by $t_p$ which is
given by the condition $ D_{A}(t_{p})D_{B}(t_{p})=|c_{3}| $, and
we can see that before the critic time, i.e., $t<t_p$, the initial
QD can remain unchanged in the time evolution while the QD decays
after the critic time, i.e., $t>t_p$. Therefore, the time $t_p$ is
the DFE time of the QD. In what follows we calculate explicitly
analytical expression of the QD in the whole time evolution to
indicate how to control the DFE time of the QD.

In the DFE regime of the QD, i.e.,  $t<t_{p}$, since $
D_{A}(t_{p})D_{B}(t_{p})>|c_{3}|$, we have $\chi
(t)=D_{A}(t)D_{B}(t)\equiv D_{A}D_{B}$. Consequently, the
classical correlation reads
\begin{eqnarray}
\mathcal{C}(\rho _{S}(t)) &=&\frac{1}{2}\Omega_{-}(t)\log
_{2}\Omega_{-}(t)+\frac{1}{2}\Omega_{+}(t)\log _{2}\Omega_{+}(t),
\nonumber \\ \label{qdA1}
\end{eqnarray}%
where we have introduced $\Omega_{\pm}(t)=1\pm D_{A}(t)D_{B}(t)$.

Subtracting the classical correlation from the quantum mutual
information $\mathcal{I}(\rho _{S})$, we find that the QD is given
by
\begin{equation}
\mathcal{D}(\rho
_{S}(t))=\sum_{n=0}^{1}\frac{1}{2}[1+(-1)^{n}c_{3}]\log
_{2}[1+(-1)^{n}c_{3}],\label{qdA2}
\end{equation}
which is independent of time. Hence, in the regime of $t<t_{p}$
the QD dynamics is decoherence-free, and the initial QD can be
preserved.

In the regime of the QD decay, i.e.,  $t>t_{p}$,  since
$D_{A}(t)D_{B}(t)<|c_{3}|$, we have $\chi (t)=\left\vert
c_{3}\right\vert$. The classical correlation is a constant
\begin{equation}
\mathcal{C}(\rho
_{S}(t))=\sum_{n=0}^{1}\frac{1}{2}[1+(-1)^{n}c_{3}]\log
_{2}[1+(-1)^{n}c_{3}], \label{qdA3}
\end{equation}
then the QD reads
\begin{eqnarray}
\mathcal{D}(\rho _{S}(t)) &=&\frac{1}{2}\Omega_{-}(t)\log
_{2}\Omega_{-}(t)+\frac{1}{2}\Omega_{+}(t)\log
_{2}\Omega_{+}(t),\nonumber \\ \label{qdA4}
\end{eqnarray}
which is a decaying function of time.

The above discussion shows that the QD exhibits a sudden change at
the critic time $t_{p}$. The QD evolution dynamics displays
different behaviors before and after the critic time. The initial
QD can be preserved in the regime of  $t<t_{p}$ while the QD
decays in the time evolution regime of $t>t_{p}$. In what follows
we discuss the dependence of the DFE time of the QD upon the
initial-state parameters, the system-reservoir coupling,  and the
reservoir temperature for the case of two equal-temperature
reservoirs and the case of two unequal-temperature reservoirs,
respectively.

\begin{figure}[htb]
\includegraphics[scale=0.55]{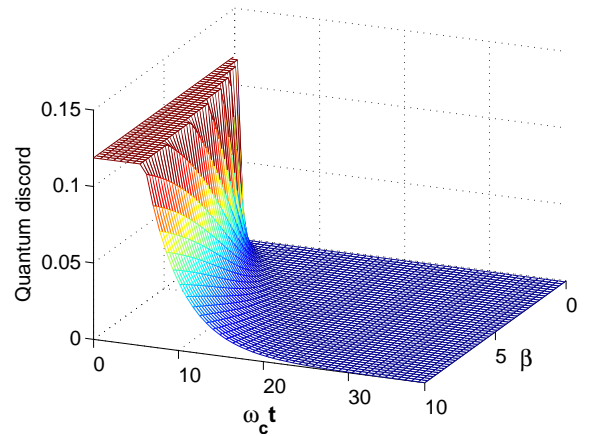}
\caption{(Color online) 3D diagram of the quantum discord as a
function of $\protect\beta $ and $\protect\omega _{c}t$  in the
case of two equal-temperature reservoirs  under the setting
$\protect\eta =0.2$, $c_{1}=1$, and $c_{2}=-c_{3}=0.4$. }
\label{fig2}
\end{figure}

\subsection{The case of two equal-temperature reservoirs}

In this subsection, we assume the two reservoirs are identical,
i.e., $\omega_{Ac}=\omega_{Bc}=\omega_{c}, \eta_A=\eta_B=\eta,
\beta_{A}=\beta_{B}=\beta$. Then we have $D_{A}(t)=D_{B}(t)=D(t)$.

\begin{figure}[htb]
\includegraphics[scale=0.55]{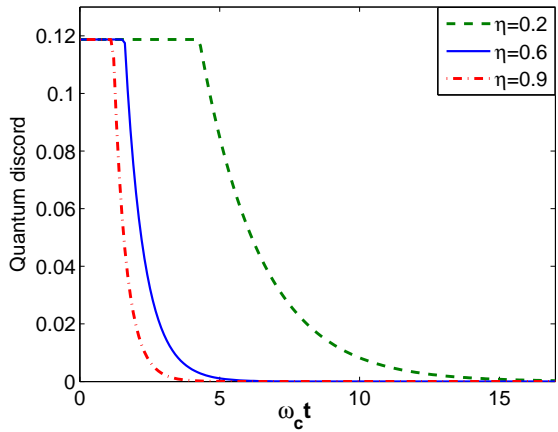}
\caption{(Color online) Plot of the quantum discord  vs the scaled
time $\protect\omega_{c}t$ in the case of two equal-temperature
reservoirs with the qubit-reservoir coupling $\protect\eta=0.2$
for dash green line, $\protect\eta=0.6$ for solid blue line,
$\protect\eta=0.9$ for dash-dot red line. Other parameters are
taken as $c_{1}=1$, $c_{2}=-c_{3}$, and $ c_3=0.4$, and
$\beta=5$.} \label{fig3.eps}
\end{figure}

At zero temperature, from Eq. (30) we can explicitly obtain the
decohering function with the form $D(t)=(1+\omega
_{c}^{2}t^{2})^{-\eta/2}$, which allows us to achieve the
expression of the QD preservation time
\begin{equation}
t_{p}=\frac{\sqrt{|c_{3}|^{-1/\eta}-1}}{\omega _{c}},
\label{qd3-19}
\end{equation}
which indicates that the DFE time depends on the initial-state
parameter $ c_{3}$, the qubit-bath coupling $\eta $, and the bath
parameter $\omega _{c}$. The DFE time increases with the decrease
of the initial-state parameter $ |c_{3}| $ or/and the decrease of
the qubit-bath coupling $\eta $. Moreover, for a set of given
parameters ($c_{3},\eta $), the lower the cut-off frequency of the
reservoir is, the longer the preservable time will be.

At finite temperatures, we can numerically study the influence of
the initial-state parameter, the qubit-bath coupling,the bath
parameter, and the bath temperature upon the QD preservation time.
In Fig. \ref{fig2}, we have plotted the QD as a function of $\beta
$ and $\omega _{c}t$.  From Fig. 2 we can see that a sudden change
of the QD occurs at the critic time of $t=t_p$, the initial QD can
be preserved in the time evolution regime of $0\le t \le t_p$.
However, The QD decays in the time evolution regime of $t>t_p$. We
can also see that the decrease of the bath temperature prolongs
the DFE time of the initial QD and slows  the decay rate of the
QD.

Fig.~\ref{fig3.eps} shows the effect of the qubit-reservoir
coupling $\eta$ on the QD dynamics. From Fig. 3 we can see that
the DFE time of the initial QD $t_{p}$ increases and the QD decay
after the time $t_{p}$ becomes slower when the system-reservoir
coupling constant $\eta $ decreases. Hence, we can conclude that
the smaller the qubit-reservoir coupling  $\eta $, the more robust
the QD against decoherence.

\begin{figure}[htb]
\includegraphics[scale=0.55]{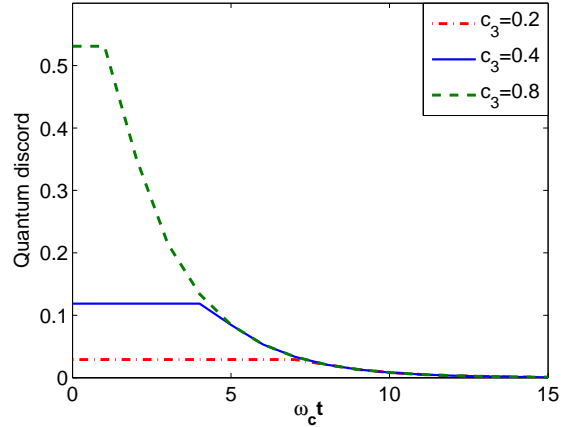}
\caption{(Color online) Plot of the quantum discord vs the scaled
time $\protect\omega_{c}t$ in the case of two equal-temperature
reservoirs with the initial-state parameter $\left\vert
c_{3}\right\vert=0.2$ for dash green line, $\left\vert
c_{3}\right\vert=0.4$ for solid blue line, and $\left\vert
c_{3}\right\vert=0.8$ for dash-dot red line. Other parameters are
taken as $\eta=0.2$, $c_{1}=1$, $c_{2}=-c_{3}$, and $\beta=5$.}
\label{fig4.eps}
\end{figure}

The effect of the initial-state parameter $\left\vert
c_{3}\right\vert $ on the QD dynamics is shown in
Fig.~\ref{fig4.eps} where we have plotted the time evolution of
the QD for $\beta=5$. From Fig. 4 we can see that one can make the
DFE time of the initial QD $t_{p}$ longer by decreasing the
magnitude of the initial-state parameter $|c_{3}|$.

\begin{figure}[htb]
\includegraphics[scale=0.4]{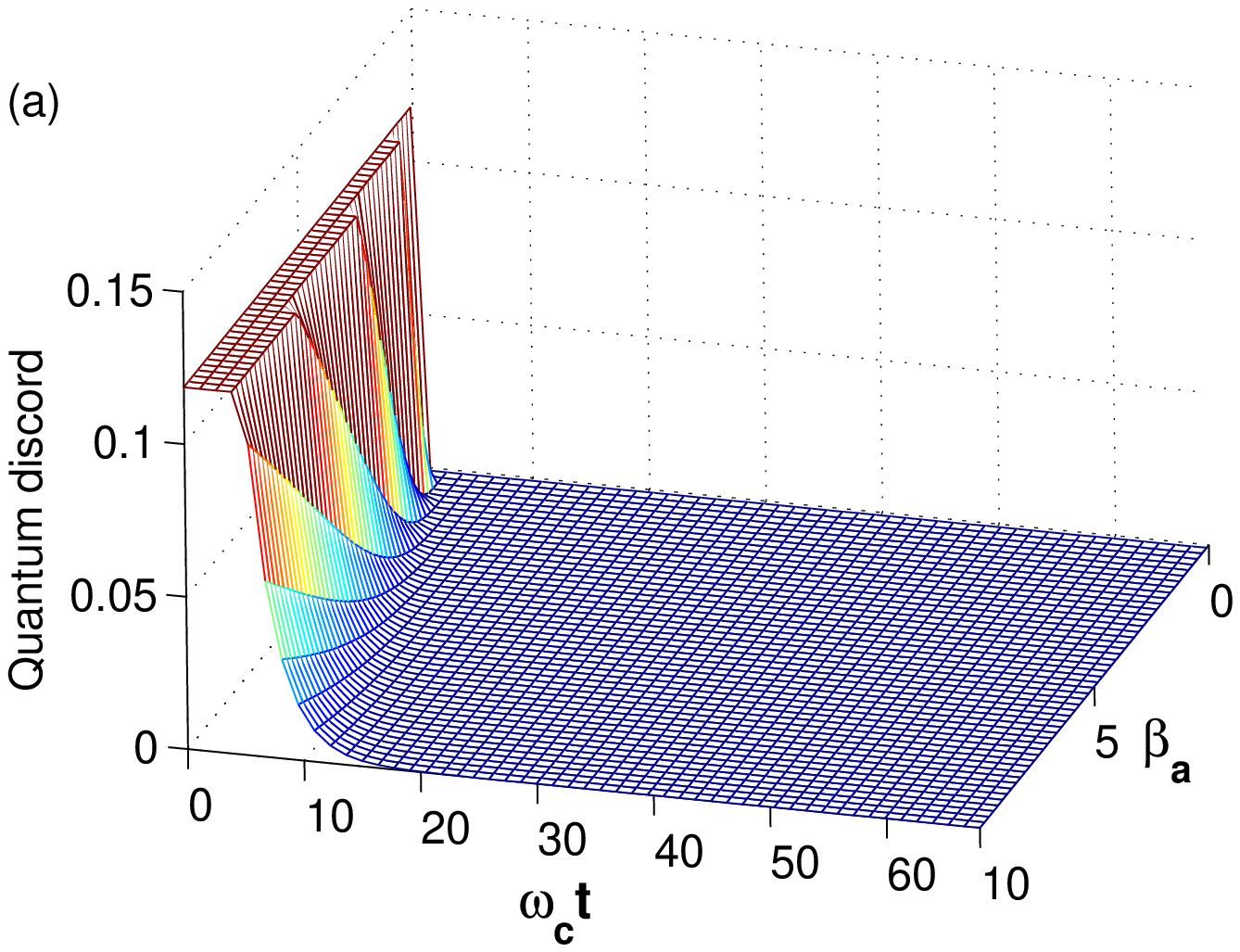}
\includegraphics[scale=0.4]{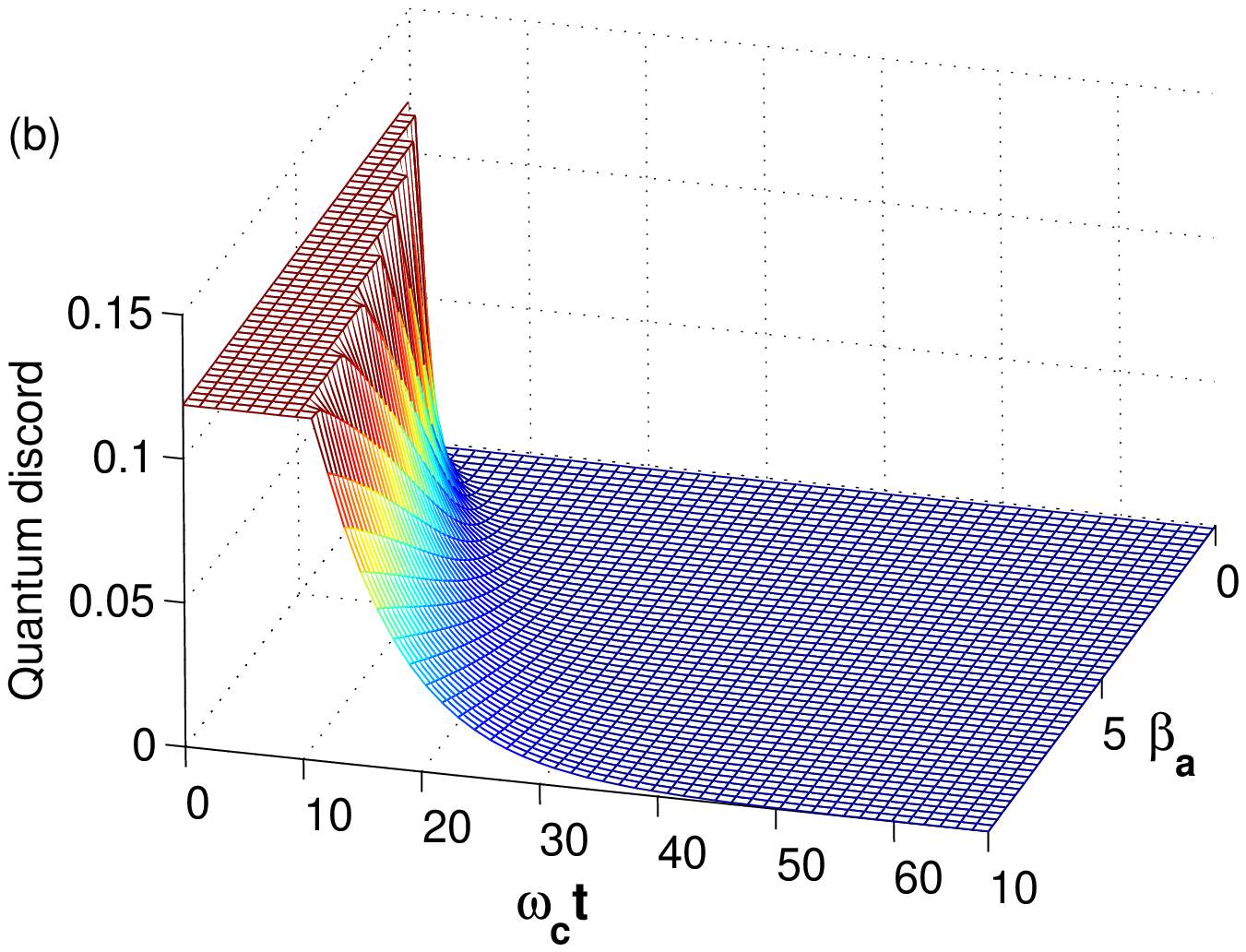}
\includegraphics[scale=0.4]{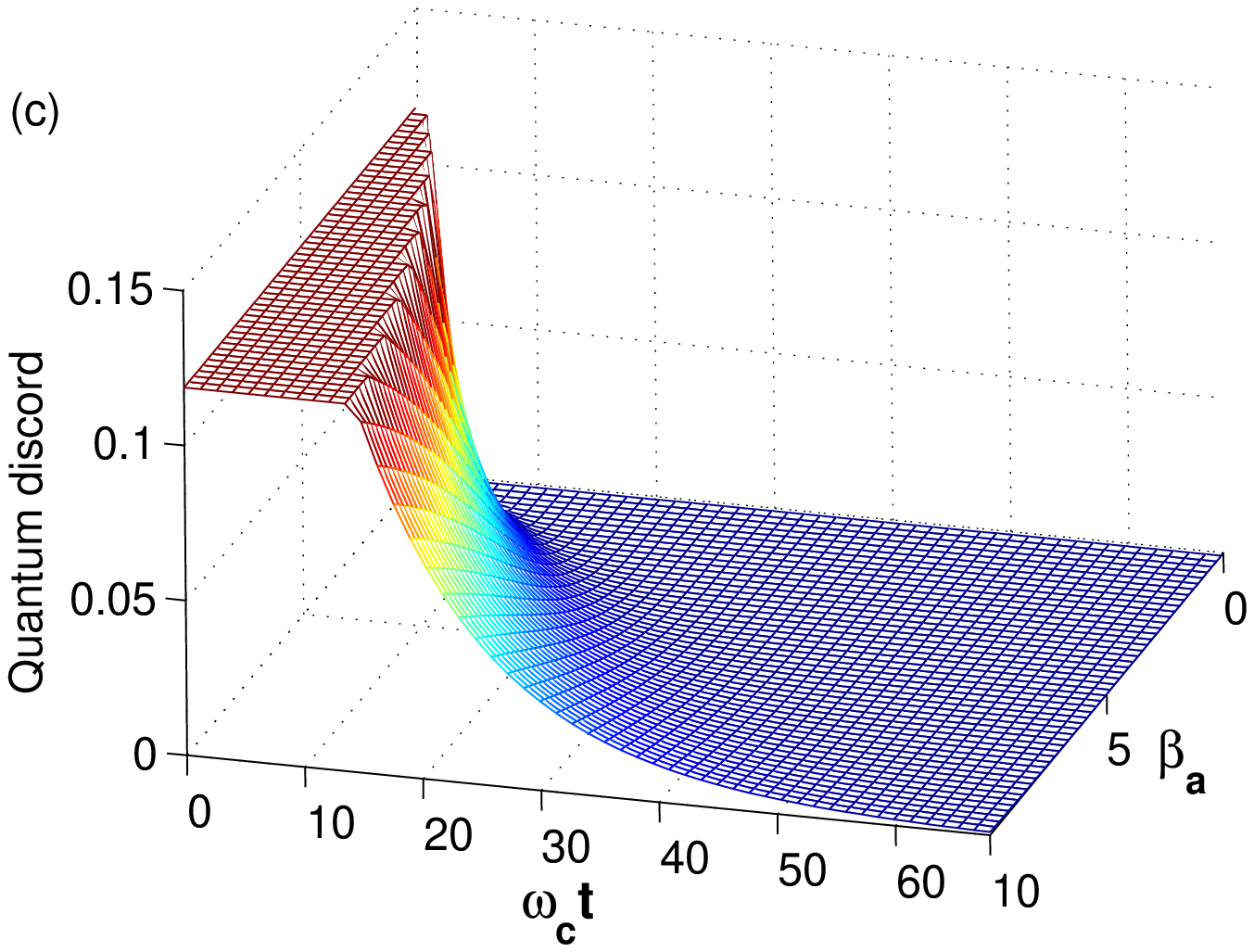}
\caption{(Color online)  3D diagram of the quantum discord as a
function of $\protect\beta_{a} $ and $\protect\omega _{c}t$ in the
case of two unequal-temperature reservoirs under the setting:
$\protect\eta =0.12$, $c_{1}=1$ and $c_{2}=-c_{3}=0.4$ for
different value of $\kappa$, (a) $\kappa=0.2$, (b) $\kappa=1$, (c)
 $\kappa=5$. } \label{fig5}
\end{figure}

\subsection{The case of two unequal-temperature reservoirs}

In this subsection, we the influence of the temperature difference
the QD dynamics  when the two reservoirs have different
temperature. We suppose that  $\omega_{Ac}=\omega_{Bc}=\omega_{c},
\eta_A=\eta_B=\eta$, and the temperature of two reservoirs
satisfies the relation $T _{a}=\kappa T _{b}$, i.e.,   $\beta
_{b}=\kappa\beta _{a}$ where $\kappa$ is a parameter to denote the
temperature difference between the two reservoirs. In case of two
unequal-temperature reservoirs, the QD dynamics is described
analytically Eqs. (33) and (35). In Fig. 5, we have plotted the
dynamic evolution of the QD with respect to $\beta_{a}$  for
different values of the temperature-difference parameter $\kappa$
when the two qubits are initially prepared in the $X$-type state
with the state parameters $c_{1}=1$, $c_{2}=-c_{3}$, and $
c_3=0.4$.

From Fig. 5 we can see that the temperature difference between the
two reservoirs does not change the maximal value of the QD but
does change the DFE time of the QD in the dynamic evolution. The
maximal value of the QD is stable for different
temperature-difference parameter $\kappa$ of the two reservoirs.
The decay rate of the QD decreases with the increase of the
temperature difference parameter $\kappa$ for the given initial
state. From Fig. 5 we can also see that the larger the temperature
difference parameter $\kappa$, the longer the DFE time $t_{p}$.
Therefore, we  conclude that the DFE time can be prolonged under
certain conditions through the increase of the temperature
difference between the two reservoirs.

\section{\label{Sec:4}Concluding remarks}
In conclusion, we have studied the DFE dynamics of the QD for two
qubits in two finite-temperature environments through
investigating the QD dynamics of the two qubits coupled to
independent reservoirs. We have chosen the initial state of the
two qubits to be $X$-type states which is known to exhibit quantum
correlations described by the QD. We have shown that reservoir
temperature significantly affects the DFE time of the QD. We have
also indicated that it is possible to control the DFE dynamics of
the two qubits and to prolong the DFE time by choosing suitable
parameters of the two-qubit system and their environments. These
results shed new light on the QD control which would be a new
direction in quantum information processing.

For $X$-type initial states with the state parameters
$c_{1}=1,c_{2}=-c_{3}$ and $|c_{3}|<1$, we have investigated the
QD dynamics of the two qubits in detail when the two independent
reservoirs are Ohmic. In the case of the two zero-temperature
reservoirs, we have shown that the DFE time increases with the
decrease of the initial-state parameter $ |c_{3}| $ or/and the
decrease of the  qubit-bath coupling $\eta $. Moreover, for a set
of given parameters ($c_{3},\eta $), the lower the cut-off
frequency of the reservoirs, the longer the DFE time is. In the
case of the two nonzero equal-temperature reservoirs, we have
indicated that the decrease of the bath temperature can prolong
the DFE time of the initial QD and slows the QD decay rate. In the
case of the two unequal-temperature reservoirs, it is found that
the temperature difference between the two reservoirs does not
change the maximal value of the QD but does change the DFE time in
the dynamic evolution. The larger the temperature difference
parameter, the longer the DFE time $t_{p}$. In this sense, we
conclude that the DFE time can be prolonged under certain
conditions through the increase of the temperature difference
between the two reservoirs.

\acknowledgments This work is supported by the Program for New
Century Excellent Talents in University (NCET-08-0682), NSFC under
Grant Nos.11075050 and 11074071, NFRPC under Grant
No.2007CB925204, the  PCSIRT under Grant No. IRT0964, the
Project-sponsored by SRF for ROCS, SEM [2010]609-5, the Key
Project of Chinese Ministry of Education (No.210150), Projects
Supported by Scientific Research Fund of Hunan Provincial
Education Department No.~09B063, No.~09C638 and No.~09C227, and
Research Fund of Hunan First Normal University No.~XYS09N07.


\vspace*{-0.1in} \bigskip





\begin{thebibliography}{99}
\bibitem{RMP73(09)00565} J.M. Raimond, M. Brune, S. Haroche, Rev. Mod. Phys. 73 (2009) 565.

\bibitem{RMP77(05)00633} M. Fleischhauer, A. Imamoglu, J.P. Marangos, Rev. Mod. Phys. 77 (2005) 633.

\bibitem{RMP73(01)00357} Y. Makhlin, G. Schoen, A. Shnirman,  Rev. Mod. Phys. 73 (2001) 357.

\bibitem{physrep04} M. Blencowe, Phys. Rep. 395 (2004) 159.

\bibitem{physrep02} M. Keyl, Phys. Rep. 369 (2002) 431.

\bibitem{physrep05} F. Mintert, A.R.R. Carvalho, M. Ku\'{s}, A. Buchleitner, Phys. Rep. 415 (2005) 207.

\bibitem{RMP81(09)00865} R. Horodecki, P. Horodecki, M. Horodecki, K. Horodecki, Rev. Mod. Phys. 81 (2009) 865.

\bibitem{JPA34} L. Henderson, V. Vedral, J. Phys. A: Math. Theor.  34 (2001) 6899.

\bibitem{oh3PRL89} J. Oppenheim, M. Horodecki, P. Horodecki, R. Horodecki, Phys. Rev. Lett. 89 (2002) 180402.

\bibitem{GPWPRA72} B. Groisman, S. Popescu, A. Winter, Phys. Rev. A 72 (2005) 032317.

\bibitem{SLPRA77} S. Luo, Phys. Rev. A 77 (2008) 022301.

\bibitem{VedPRL104} K. Modi, T. Paterek, W. Son, V. Vedral, M. Williamson, Phys. Rev. Lett. 104 (2010) 080501.

\bibitem{DLZPRL101} D.L. Zhou, Phys. Rev. Lett. 101 (2008) 180505.

\bibitem{zerekdMD} W.H. Zurek, Phys. Rev. A 67 (2003) 012320.

\bibitem{PRL105(10)020503} P. Giorda, M.G.A.  Paris, Phys. Rev. Lett. 105 (2010) 020503.

\bibitem{PRL100(08)090502} M. Piani, P. Horodecki, R. Horodecki, Phys. Rev. Lett. 100 (2008) 090502\\
M. Piani, M. Christandl, C.E. Mora, P. Horodecki, ibid. 102 (2009)
250503.

\bibitem{JPA41(08)205301} C. A. Rodriguez-Rosario, K. Modi, A. Kuah, A. Shaji, E.C.G. Sudarshan,  J. Phys. A: Math. Theor. 41 (2008) 205301.

\bibitem{Lidar102} A. Shabani, D.A. Lidar,  Phys. Rev. Lett. 102 (2009) 100402.

\bibitem{DGPRA79} A. Datta, S. Gharibian, Phys. Rev. A 79 (2009) 042325.

\bibitem{Yuan43} J.B. Yuan, L.M. Kuang, J.Q.  Liao, J. Phys. B: At. Mol. Opt. Phys. 43 (2010) 165503.

\bibitem{zurekPRL88} H. Ollivier, W.H. Zurek, Phys. Rev. Lett. 88 (2010) 017901.

\bibitem{Datta100} A. Datta, A. Shaji, C.M. Caves, Phys. Rev. Lett. 100 (2008) 050502.

\bibitem{LBAW101} B.P. Lanyon, M. Barbieri, M.P. Almeida, A.G. White, Phys. Rev. Lett. 101 (2008) 200501.

\bibitem{MWPRA81} J. Maziero, T. Werlang, F.F. Fanchini, L.C. C\'{e}leri, R.M. Serra,Phys. Rev. A 81 (2010) 022116.

\bibitem{Wang2009}  B. Wang, Z.Y. Xu, Z.Q. Chen, M. Feng,  Phys. Rev. A 81 (2010) 014101.

\bibitem{Piilo2010}   L. Mazzola, J.  Piilo, S. Maniscalco, Phys. Rev. Lett. 104 (2010) 200401.

\bibitem{VedPRA80} J. Maziero, L.C. C\'{e}leri, R.M. Serra, V. Vedral, Phys. Rev. A 80 (2009) 044102.


\bibitem{cel}  L.C. C\'{e}leri, A.G.S. Landulfo, R.M. Serra, G.E.A. Matsas, Phys. Rev. A 81 (2010) 062130.

\bibitem{xu1} J.S. Xu, C.F. Li, C.J. Zhang, X.Y. Xu, Y.S. Zhang, G.C. Guo, Phys. Rev. A 82 (10) 042328.

\bibitem{xu2} J.S. Xu, X.Y. Xu, C.F. Li, C.J. Zhang, X.B. Zou, G.C. Guo, Nat. Commun. 1 (2010) 7.

\bibitem{WSPRA80} T. Werlang, S. Souza, F.F. Fanchini, C.J. Villas Boas, Phys. Rev. A 80 (2009) 024103.

\bibitem{FWPRA81} F.F. Fanchini, T. Werlang, C.A. Brasil, L.G.E. Arruda, A.O. Caldeira, Phys. Rev. A 81 (2010) 052107.

\bibitem{FAPRA81} A. Ferraro, L. Aolita, D. Cavalcanti, F.M. Cucchietti, A. Ac\'{\i}n, Phys. Rev. A 81 (2010) 052318.

\bibitem{Yu2004}  T. Yu, J.H.  Eberly, Phys. Rev. Lett. 93 (2004) 140404; T. Yu, J.H.  Eberly, Science 323 (2009) 598.

\bibitem{Bellomo2007} B. Bellomo, R.L. Franco, G. Compagno, Phys. Rev. Lett. 99 (2007) 160502.

\bibitem{Choi2007}    J. Laurat, K.S.  Choi, H. Deng, C.W. Chou, H.J. Kimble, Phys. Rev. Lett. 99 (2007) 180504.

\bibitem{opez2008}    C.E. L\'{o}pez, G. Romero, F. Lastra, E. Solano, J.C. Retamal, Phys. Rev. Lett. 101 (2008) 080503.

\bibitem{Qasimi2008}  A. Al-Qasimi, D.F.V. James, Phys. Rev. A 77 (2008) 012117

\bibitem{Maniscalco2008} S. Maniscalco, F. Francica, R.L. Zaffino, N.L. LoGullo, F. Plastina, Phys. Rev. Lett. 100 (2008) 090503.


\bibitem{BBcontrol} L. Viola, S. Lloyd, Phys. Rev. A 58 (1998) 2733.

\bibitem{Kuang1999} L.M. Kuang, H.S. Zeng, Z.Y. Tong, Phys. Rev. A 60 (1999) 3815; L.M. Kuang, Z.Y. Tong, Z.W. Ouyang, H.S. Zeng, hys. Rev. A 61 (1999) 013608.

\bibitem{VedPRL90} V. Vedral, Phys. Rev. Lett. 90 (2003) 050401.

\bibitem{LegRMP59} S. Chakravarty, A.J. Leggett, Phys. Rev. Lett. 52 (1984) 5; A.J. Leggett, S. Chakravarty, A.T. Dorsey, M.P.A. Fisher, A.
Garg, W. Zwerger, Rev. Mod. Phys. 59 (1987) 1.




\end{thebibliography}
\end{document}